\documentstyle[12pt,epsf]{article}

\begin{document}

\begin{titlepage}

\null
\begin{flushright}
CERN-TH.7404/94\\
ILL-(TH)-94-20
\end{flushright}
\vspace{20mm}

\begin{center}
\bf\Large Entropy and the approach to the \\
          thermodynamic limit\\
          in three-dimensional simplicial gravity
\end{center}

\vspace{5mm}

\begin{center}
{\bf S. Catterall\footnote{Permanent address: Physics Department, Syracuse
University, Syracuse, NY 13244}}\\
TH-Division, CERN CH-1211,\\
Geneva 23, Switzerland.\\
{\bf J. Kogut}\\
Loomis Laboratory, University of Illinois at Urbana,\\
1110 W. Green St, Urbana, IL 61801.\\
{\bf R. Renken}\\
Department of Physics, University of Central Florida,\\
Orlando, FL 32816.
\end{center}

\begin{center}
\today
\end{center}

\vspace{10mm}
      
\begin{abstract}
We present numerical results supporting the existence of an exponential
bound in the dynamical triangulation model of three-dimensional 
quantum gravity. Both the critical coupling and various other
quantities show a slow power law approach to the infinite volume limit. 
\end{abstract}

\vfill
\begin{flushleft}
CERN-TH.7404/94\\
August 1994
\end{flushleft}

\end{titlepage}

\section*{Introduction}

Much interest has been generated recently in lattice models for euclidean
quantum gravity based on dynamical triangulations 
\cite{mig3,amb3,gross,mig4,amb4,brug4,us4,smit4}.
The study of these models was prompted by the success
of the same approach in the case of two dimensions, see for example 
\cite{david}.
The primary input to these models is the ansatz that the partition
function describing the fluctuations of a continuum geometry can be
approximated by performing a weighted sum over all simplicial manifolds
or triangulations $T$.

\begin{equation}
Z=\sum_{T}\rho\left(T\right)
\label{eqn1}
\end{equation}

In all the work conducted so far the topology of the lattice has been
restricted to the sphere $S^d$. The weight function $\rho\left(T\right)$
is taken to be of the form

\begin{equation}
\rho\left(T\right)=e^{-\kappa_d N_d +\kappa_0 N_0}
\end{equation}

The coupling $\kappa_d$ represents a bare lattice cosmological constant
conjugate to the total volume (number of $d$-simplices $N_d$) whilst
$\kappa_0$ plays the role of a bare Newton constant coupled to the total
number of nodes $N_0$.

We can rewrite eqn. \ref{eqn1} by introducing the entropy function
$\Omega_d\left(N_d, \kappa_0\right)$ which counts the number of triangulations
with volume $N_d$ weighted by the node term. This the primary
object of interest in this note.

\begin{equation}
Z=\sum_{N_d} \Omega_d\left(N_d, \kappa_0\right) e^{-\kappa_d N_d}
\end{equation}

For this partition sum to exist it is crucial that the entropy function
$\Omega_d$ increase no faster than exponentially with volume. For two
dimensions this is known \cite{2dbound} but until recently the 
only evidence for this in higher dimensions came from numerical simulation. 
Indeed for
four dimensions there is some uncertainty in the status of this bound
\cite{usbound,ambbound,smit4}. 

With this in mind we have conducted
a high statistics study of the three dimensional model
at $\kappa_0=0$, extending the
simulations reported in \cite{varsted} by an order of magnitude in 
lattice volume. While in the course of preparing this manuscript
we received a paper \cite{boulatov} in which an argument for the bound in
three dimensions is given. Whilst we observe a rather slow
approach to the asymptotic, large volume limit, our results are
entirely consistent with the existence of such a bound. However, the
predicted variation of the mean node number with volume is not seen, rather
the data supports a rather slow power law approach to the infinite volume
limit.

If we write $\Omega_3\left(N_3\right)$ as

\begin{equation}
\Omega_3\left(N_3\right)=ae^{\kappa_3^c\left(N_3\right) N_3}
\end{equation} 
the effective critical cosmological constant $\kappa_3^c$ is
taken dependent on the volume and a bound implies that $\kappa_3^c\to
{\rm const}
 < \infty$
as $N_3\to\infty$. In contrast for a model where the entropy
grew more rapidly than exponentially $\kappa_3^c$ would
diverge in the thermodynamic limit. 

To control the volume 
fluctuations we add a further
term to the action of the form $\delta S=\gamma\left(N_3-V\right)^2$.
Lattices with $N_3\sim V$ are distributed according to the correct
Boltzmann weight up to correction terms of order $O\left(1\over \sqrt{\gamma}V
\right)$ where we use $\gamma=0.005$ in all our runs. This error
is much smaller than our statistical errors and can hence be
neglected.

Likewise, as a first approximation, we can set $\kappa_3^c$ equal to
its value at the mean of the volume distribution $V$ which allows us to
compute the expectation value of the volume exactly since the
resultant integral is now a simple gaussian. We obtain

\begin{equation}
\left\langle N_3\right\rangle = {1\over 2\gamma}\left(\kappa_3^3\left(V\right)
-\kappa_3\right)+V
\label{eqn2}
\end{equation}

Equally, by measuring the mean volume $\left\langle N_3\right\rangle$ 
for a given input value of
the coupling $\kappa_3$ we can estimate $\kappa_3^c\left(V\right)$ for
a set of mean volumes $V$.
The algorithm we use to generate a Monte Carlo
sample of three dimensional lattices is described in \cite{simon}.
We have simulated systems with volumes up to $128000$ 3-simplices and
using up to $400000$ MC sweeps (a sweep is defined as $V$ {\it attempted}
elementary updates of the triangulation where $V$ is the
average volume). 

\begin{figure}
\begin{center}
\leavevmode
\epsfxsize=400pt
\epsfbox{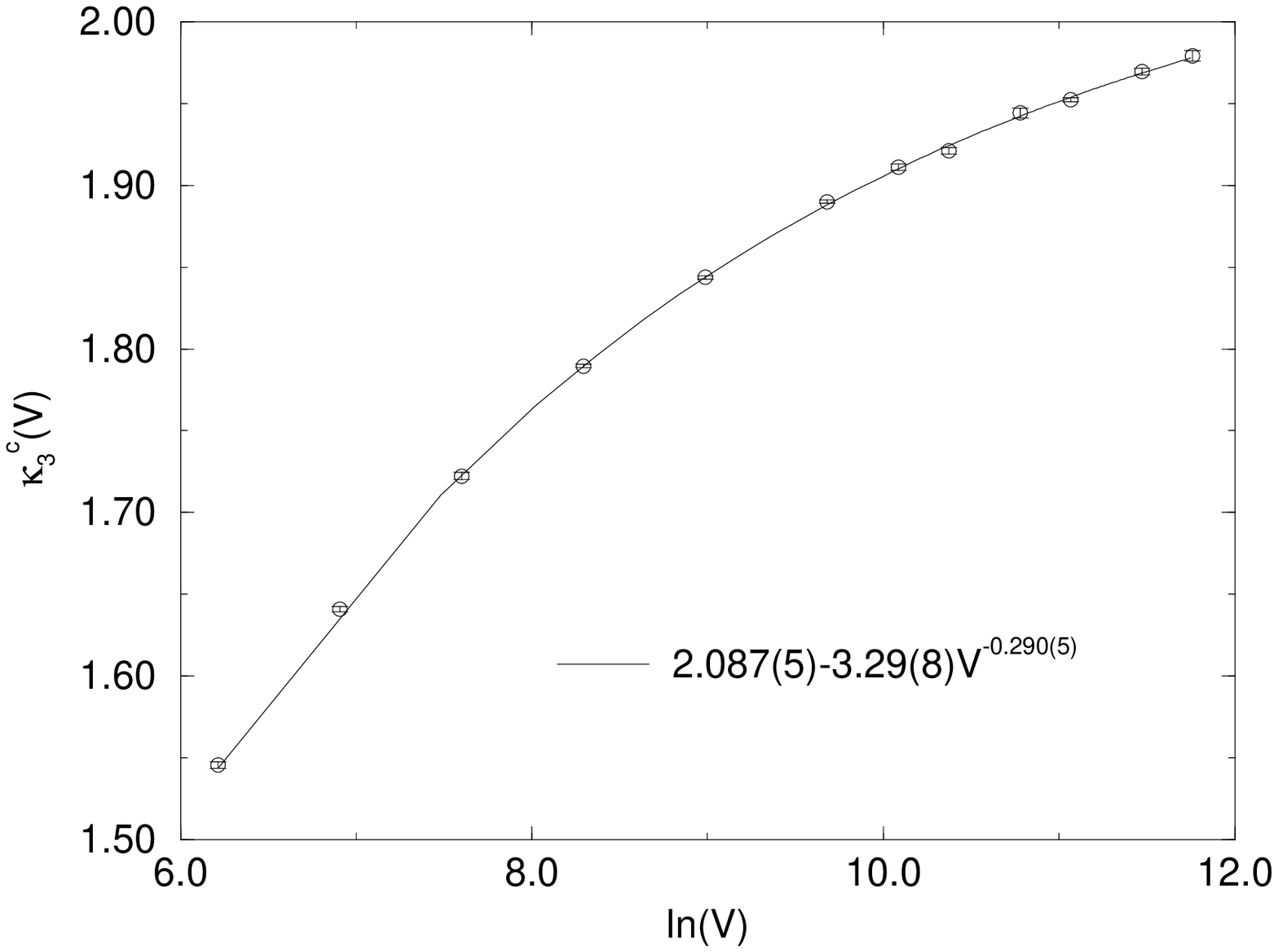}
\caption{Critical coupling vs volume }
\label{fig1}
\end{center}
\end{figure}

Our results for $\kappa_3^c\left(V\right)$, computed this way,
are shown in fig. \ref{fig1} as a function of $\ln V$. The choice of
the latter scale is particularly apt as the presence of a 
factorial growth in $\Omega_3$ would be signaled by a logarithmic
component to the effective $\kappa_3^c\left(V\right)$. As the plot
indicates there is no evidence for this. Indeed, the best fit
we could make corresponds to a {\it convergent} power law 

\begin{equation}
\kappa_3^c\left(V\right)=\kappa_3^c\left(\infty\right) + a V^{-\delta}
\end{equation}

If we fit all of our data we obtain best fit parameters 
$\kappa_3^c\left(\infty\right)=2.087(5)$,
$a=-3.29(8)$
and $\delta=0.290(5)$ 
with a corresponding $\chi^2$ per degree of
freedom $\chi^2=1.3$ at $22\%$ confidence (solid line shown). 
Leaving off the
smallest lattice $V=500$ yields a statistically consistent fit with
an even better $\chi^2=1.1$ at $38\%$ confidence. 
We have further tested the stability
of this fit by dropping either the small volume data ($V=500-2000$
inclusive), the large volume data ($V=64000-128000$ inclusive) or
intermediate volumes ($V=8000-24000$). In each of these cases 
the fits were good and yielded
fit parameters consistent with our quoted best fit to all the data.
Furthermore, these numbers
are consistent with the earlier study \cite{varsted}. We are
thus confident that this power law is empirically a very
reasonable parameterisation of the approach to the 
thermodynamic limit. Certainly, our conclusions must be that
the numerical data {\it strongly} favour the existence of a 
bound.

One might object that the formula used to compute $\kappa_3^c$ is only
approximate (we have neglected the variation of
the critical coupling over the range of fluctuation of
the volumes). This, in turn might yield finite volume corrections
which are misleading. To check for this we have extracted $\kappa_3^c$
directly from the measured distribution of 3-volumes $Q\left(N_3\right)$.
To do this we computed a new histogram $P\left(N_3\right)$

\begin{equation}
P\left(N_3\right)=Q\left(N_3\right)e^{\kappa_3 N_3+\gamma\left(N_3-V\right)^2}
\end{equation}

\begin{figure}
\begin{center}
\leavevmode
\epsfxsize=400pt
\epsfbox{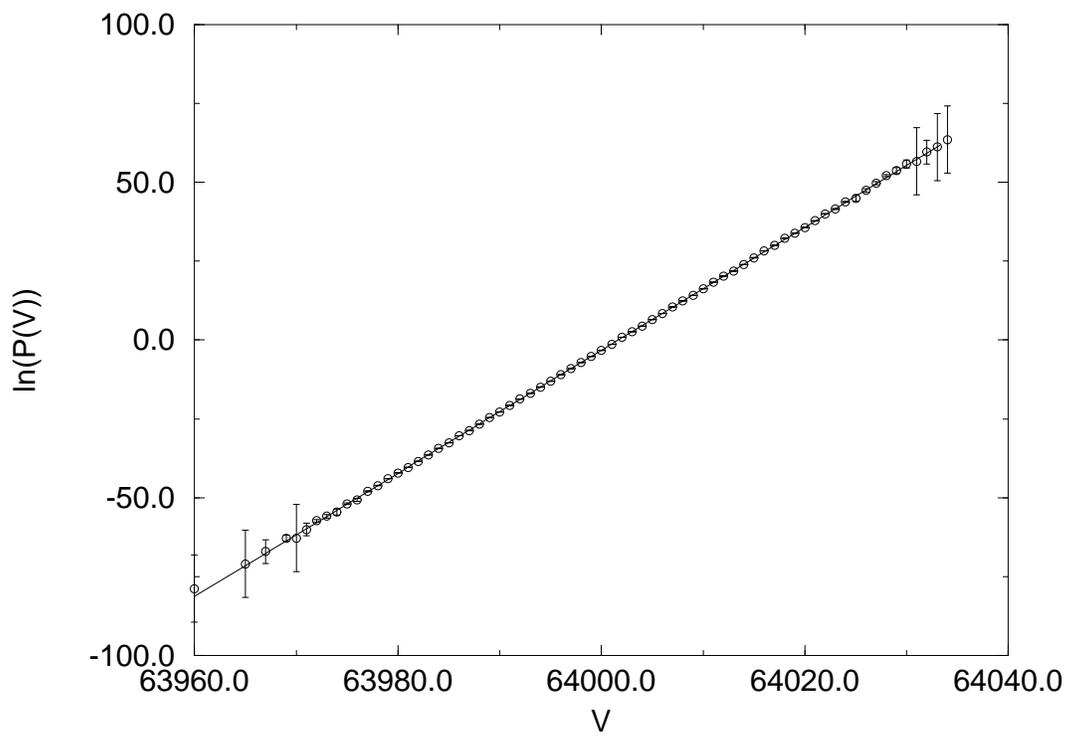}
\caption{Modified distribution of 3-volumes}
\label{fig2}
\end{center}
\end{figure}

As an example we show in fig. \ref{fig2} the logarithm of this 
quantity as a function of volume
for $V=64000$.
The gradient of the straight line fit shown is an unbiased estimator of
the critical coupling $\kappa_3^c\left(64000\right)$. The value
of $1.9516(10)$ compares very favourably with the value $\kappa_3^c\left(
64000\right)=1.9522(12)$ obtained 
using eqn. \ref{eqn2}.
Indeed, this might have
been anticipated since we might expect corrections
to eqn. \ref{eqn2} to be of magnitude 
$O\left(V^{-\left(1+\delta\right)}\right)$ which even for the smallest
volumes used in this study is again much smaller than our statistical
errors.

In addition to supplying a proof of the
exponential bound in \cite{boulatov} Boulatov also 
conjectures a relation between the
mean node number and volume in the crumpled phase of the model (which
includes our node coupling $\kappa_0=0$). This has the form

\begin{equation}
\left\langle N_0/V\right\rangle=c1 + c2 {\ln(V)\over V}
\label{eqn3}
\end{equation}

\begin{figure}
\begin{center}
\leavevmode
\epsfxsize=400pt
\epsfbox{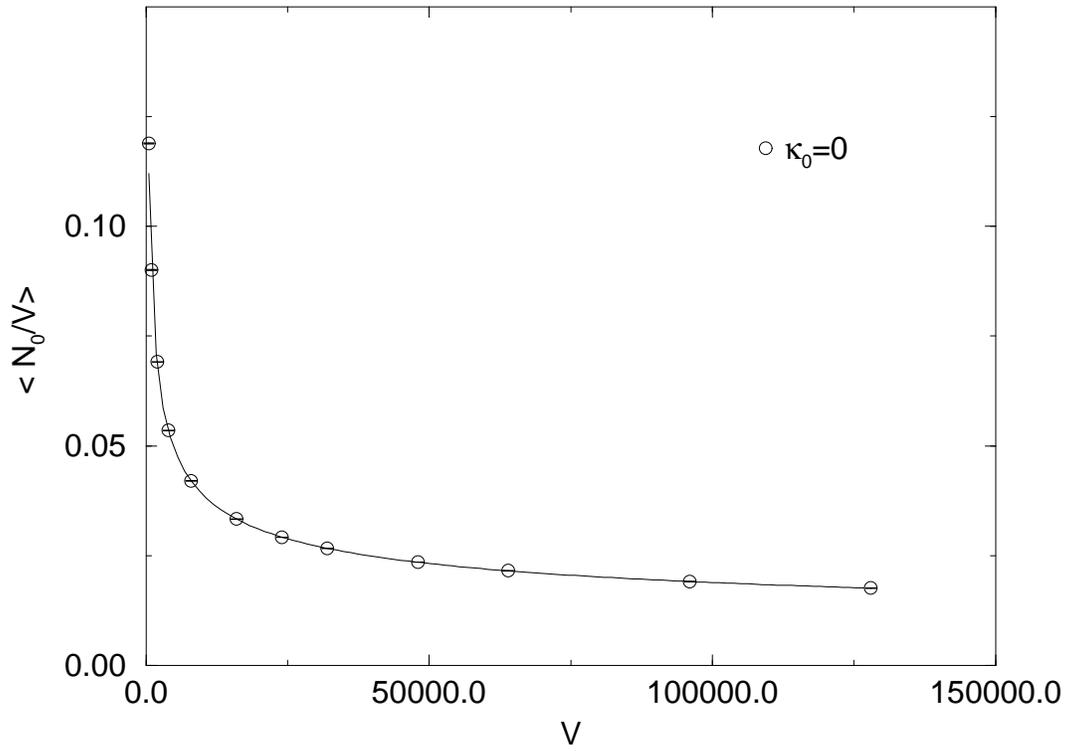}
\caption{Number of nodes per unit volume}
\label{fig3}
\end{center}
\end{figure}

Our data for this quantity are shown in fig. \ref{fig3}. Whilst it
appears that the mean coordination may indeed plateau for large volumes the
approach to this limit seems not to be governed by the corrections
envisaged in eqn. \ref{eqn3} -- it is simply impossible to fit the 
results of the simulation with this functional form. Indeed, the best
fit we could obtain corresponds again to a simple converging power with small
exponent $\left\langle N_0/V\right\rangle \sim b + cV^{-d}$. The fit
shown corresponds to using all lattices with volume $V\ge 8000$ and
yields $b=0.0045(1)$, $c=1.14(2)$ and power $d=0.380(3)$ ($\chi^2=1.6$). 
Fits to subsets of the large volume data yield consistent
results.

\begin{figure}
\begin{center}
\leavevmode
\epsfxsize=400pt
\epsfbox{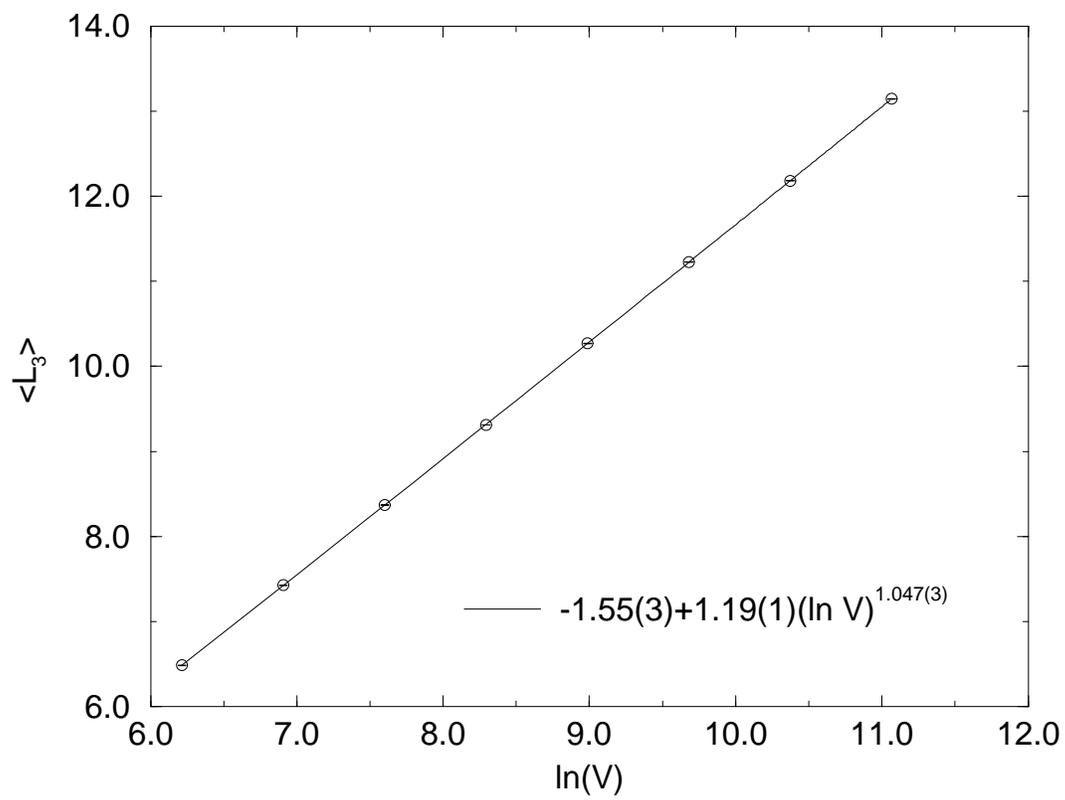}
\caption{Mean intrinsic extent}
\label{fig4}
\end{center}
\end{figure}

Finally, we show in fig. \ref{fig4}, a plot of the mean intrinsic size
of the ensemble of simplicial graphs versus their volume. This
quantity is just the average geodesic distance
(in units where the edge lengths are all unity) between two
randomly picked sites. The solid line
is an empirical fit of the form

\begin{equation}
L_3=e + f\left(\ln{V}\right)^{g}
\end{equation}

Clearly, the behaviour is close to logarithmic (as appears also to
be the case in four dimensions \cite{us4}), the exponent $g=1.047(3)$ from
fitting all the data ($\chi^2=1.7$ per degree of freedom). 
This
is indicative of the crumpled nature of the typical simplicial
manifolds dominating the partition function at this node coupling.
It is tempting to speculate that the true behaviour is simply logarithmic
and the deviation we are seeing is due to residual finite volume effects.

Alternatively, we can fit the data as a linear combination of the form 

\begin{equation}
L_3=e + f\ln{V} +g\ln{\ln{V}}
\end{equation}

This gives a competitive fit with $e=-1.45(4)$, $f=1.438(4)$ and
$g=-0.55(3)$ with $\chi^2=1.6$. One might be tempted to favour this
fit on the grounds that it avoids the problem of a power close to but
distinct from unity. However, the situation must remain
ambiguous without further theoretical insight. 

To summarise this brief note we have obtained numerical results
consistent with the existence of an exponential bound in a 
dynamical triangulation model of three dimensional quantum gravity.
Thus, practical numerical studies can reveal the bound argued for
in \cite{boulatov}. Our results also favour the existence of a finite
(albeit large $\sim 200$)
mean coordination number in the infinite volume limit in the crumpled phase.
However, the nature of the finite volume corrections to the latter appear
very different from those proposed in \cite{boulatov}. Indeed, both for
the critical coupling and mean coordination number we observe large
power law corrections with small exponent.
Finally, we show data for the scaling of the mean intrinsic extent with
volume which suggests a very large (possible infinite) fractal
dimension for the typical simplicial manifolds studied.
 
This work was supported, in part, by NSF grant PHY 92-00148. Some
calculations were performed on the Florida State University Cray YMP.

\vfill
\newpage

\end{document}